\documentclass[aps,pre,english,reprint,twocolumn,showpacs,superscriptaddress,groupedaddress]{revtex4}
\usepackage[LGR,T1]{fontenc}
\usepackage[latin9]{inputenc}
\usepackage{verbatim}
\usepackage{textcomp}
\usepackage{amstext}
\usepackage{graphicx}
\usepackage{xfrac}
\makeatletter

\DeclareFontEncoding{LGR}{}{}
\DeclareTextSymbol{\~}{LGR}{126}

\@ifundefined{textcolor}{}
{%
 \definecolor{BLACK}{gray}{0}
 \definecolor{WHITE}{gray}{1}
 \definecolor{RED}{rgb}{1,0,0}
 \definecolor{GREEN}{rgb}{0,1,0}
 \definecolor{BLUE}{rgb}{0,0,1}
 \definecolor{CYAN}{cmyk}{1,0,0,0}
 \definecolor{MAGENTA}{cmyk}{0,1,0,0}
 \definecolor{YELLOW}{cmyk}{0,0,1,0}
}

\makeatother

\usepackage{babel}
\begin{document}

\title{Violent relaxation in two-dimensional flows with varying interaction range}

\author{A. Venaille
}

\author{ T. Dauxois}

\affiliation{Laboratoire de Physique, \'Ecole Normale Sup\'erieure de Lyon, Universit\'e de Lyon, CNRS, 46 All\'ee d'Italie, F-69364 Lyon, cedex 07,France}

\author{ S. Ruffo}

\affiliation{Dipartimento di Fisica e Astronomia and CSDC, Universit\`a di Firenze, INFN and CNISM, via G. Sansone, 1 50019 Sesto Fiorentino, Italy}

\begin{abstract}
Understanding the relaxation of a system towards equilibrium is a longstanding problem in statistical mechanics. Here we address the role of long-range interactions in this process by considering a class of two-dimensional flows where the interaction between fluid particles varies with the distance  as $\sim r^{\alpha-2}$ for $\alpha>0$. 
 {We find that  changing $\alpha$ with a prescribed initial state leads to different flow patterns}: for small $\alpha$,  a coarsening process leads to the formation of a sharp interface between two regions of homogenized $\alpha$-vorticity; for large $\alpha$, the flow is attracted to a stable dipolar structure through a filamentation process.    {Assuming that the energy $E$ and the enstrophy $Z$ are injected at a typical scale smaller than the domain scale $L$, we argue that convergence towards the equilibrium state is expected when the parameter  $\left(\frac{2\pi}{L}\right)^\alpha \frac{E}{Z}$ tends to one, while convergence towards a dipolar state occurs  systematically when this parameter tends to zero. This suggests that weak long-range interacting systems are more prone to relax towards an equilibrium state than strong long-range interacting systems.} 
\end{abstract}

\maketitle

Self-gravitating systems, non-neutral plasma and two-dimensional (2D) flows are examples of long-range interacting systems, for which the two-body
potential  decays with the interparticle distance with an exponent smaller than the dimensionality of the embedding space~\cite{longrangebook}. 
Those systems share the property to self-organize spontaneously into large scale coherent structures such as globular clusters and elliptical galaxies in astrophysics or vortices and jets in geophysics~\cite{chavanis2002statistical}. 
Equilibrium statistical mechanics provides an explanation and a prediction for this phenomenon as the most probable result of mixing in phase space. 
It allows to reduce the study of the large scale organization to a few parameters, without describing the full complexity of the dynamics
involving a huge number of degrees of freedom. 
The original idea to use statistical mechanics arguments to describe self-organization of 2D flows comes from L. Onsager himself in the framework of point vortex models~\cite{Onsager:1949_Meca_Stat_Points_Vortex,eyink2006onsager}. 
A  statistical mechanics theory for the continuous Euler dynamics has been proposed by Miller, Robert, Sommeria~\cite{Robert:1990_CRAS,Miller:1990_PRL_Meca_Stat,miller1992statistical,SommeriaRobert:1991_JFM_meca_Stat} (MRS hereafter), which has led to several successful applications to geophysical flows~\cite{Sommeria_2001_CoursLesHouches,Majda_Wang_Book_Geophysique_Stat,BouchetVenaillePR12,lucarini2013mathematical}.
This theory is the equivalent of Lynden Bell's  statistical mechanics of the Vlasov dynamics~\cite{lynden1967statistical}, which has been proven useful to address  self-organisation in self-gravitating systems~\cite{joyce2011quasistationary}, or in toy models of long-range interacting particles~\cite{antoniazzi2007exploring}.
\begin{figure}
\includegraphics[width=\columnwidth]{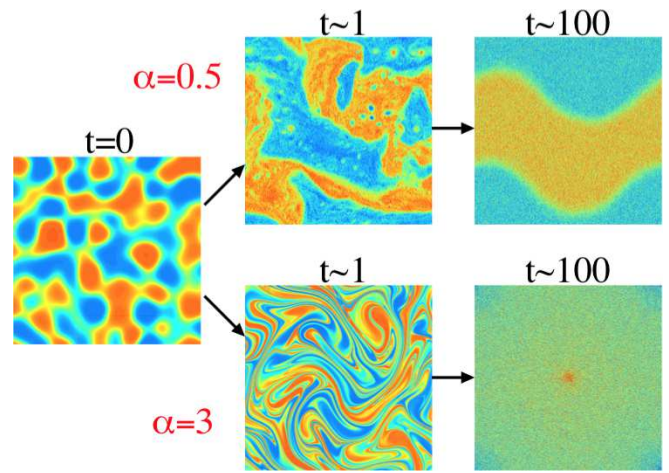}
\caption{(Color online) Temporal evolution of the $\alpha$-vorticity field during the violent relaxation of an unstable initial condition.}
\label{fig1}
\end{figure}
The equilibrium theory of long-range interacting systems is now fairly well understood~\cite{longrangebook}, but determining how and when an initially unstable condition far from equilibrium  actually relaxes towards an equilibrium state after a "violent relaxation" process~\cite{lynden1967statistical} remains a challenging problem~\cite{levin2014nonequilibrium}. 
 {Here we will focus on physical models that are continuous dynamical systems involving an infinite number of degrees of freedom, but their numerical implementation requires a discrete approximation of the dynamics.  
The "violent relaxation" corresponds to the fast evolution of the discretized system towards a stable state of the continuous dynamics, usually called quasi-stationary state (QSS). This relaxation occurs on a typical time scale independent of the discretization~\cite{yamaguchi2004,barre2006}.} 
%
%

%
One key difficulty concerning the continuous dynamics of long-range interacting systems is that  the set of stable states is usually much larger than the set of equilibrium states, and there exists other invariant measures than those predicted by the equilibrium theory~\cite{BouchetCorvellec2010}. 
Consequently, a class of unstable initial conditions may be attracted towards stable states different from the most probable one. 
Indeed, various examples of ergodicity breaking in non neutral plasma and self-gravitating systems have been reported~\cite{levin2014nonequilibrium}.
%
%
Similarly, applications of  MRS theory to freely evolving flows have led to mitigated results~\cite{tabeling2002two,YinMontgomeryClercx,morita2011collective,marston2014hyperviscosity,dritschel2015two}.
%
%

In all those previous studies, the initial conditions was varied but the range of the interaction was fixed. 
%
Here we address the role of long-range interactions during the violent relaxation of a given unstable initial condition towards QSS by considering a class of 2D flow models introduced in Ref.~\cite{PierrehumbertHeldSwanson_Chaos94}, in which interactions between fluid particles (infinitesimal fluid volumes) are labelled by a parameter $\alpha>0$. 
It includes 2D Euler dynamics ($\alpha=2$), surface quasi-geostrophic dynamics ($\alpha=1$), which is relevant to describe some aspects of atmospheric and oceanic turbulence~\cite{HeldSQG95JFM}, and a model for mantle convection ($\alpha=3$)~\cite{tran2004,weinstein1989}.
The initial goal for studying this model was to address the locality hypothesis for turbulent cascade~\cite{PierrehumbertHeldSwanson_Chaos94,smith2002turbulent,burgess2013spectral}. 
It has then been proven useful to investigate the possible emergence of finite time singularities~\cite{scott2014numerical} and conformal invariance~\cite{bernard2007inverse}.
We will see that it also sheds new light on the dynamical effects underpinning self-organisation of long-range interacting systems.\\
%
%
%

The {\em{continuous}} dynamics is expressed as the advection of the $\alpha$-vorticity $q(\mathbf{r},t)$ by a 2D incompressible velocity field $\mbox{\ensuremath{\mathbf{v}}}=(-\partial_{y}\psi,\ \partial_{x}\psi)$, with $\mathbf{r}=(x,y)$ and $\psi$ the stream function:
\begin{equation}
\partial_{t}q+ \mathbf{v}\cdot \nabla q=0,\quad q=-\left(-\Delta\right)^{\alpha/2}\psi \ ,\label{eq:continuous_dynamics}
\end{equation}
where $-\left(-\Delta\right)^{\alpha/2}$ is the fractional Laplacian defined in term of the Fourier components $q_{\mathbf{k}}=-\left|\mathbf{k}\right|^{{\alpha}}\psi_{\mathbf{k}}$ in the case a doubly periodic domain $\mathcal{D}=\left[0,\ L\right]^2$, with $\mathbf{k}=(k,\ell)$ the wave vector. Length unit has been chosen such that the domain scale is $L=2 \pi$.
The dynamics conserves the $\alpha$-energy $\mathcal{E}\left[q\right]\equiv-\int_{\mathcal{D}}\mathrm{d} \mathbf{r} \ q\psi$  and  the Casimir functionals $\mathcal{C}_{f}\left[q\right]\equiv\int_{\mathcal{D}}\mathrm{d} \mathbf{r} \ f(q),$ where $f$ is any sufficiently smooth function on $\mathcal{D}$, which includes the $\alpha$-enstrophy  $\mathcal{Z}\left[q\right]\equiv \int_{\mathcal{D}}\mathrm{d} \mathbf{r}\ q^{2}$.

The $\alpha-$energy  can be formally written as a potential energy $\mathcal{E} =\int_{\mathcal{D}}\mathrm{d}\mathbf{r}\ \int_{\mathcal{D}}\mathrm{d}\mathbf{r}^{\prime}\ q(\mathbf{r})V(\mathbf{r},\mathbf{r}^{\prime})q(\mathbf{r}^{\prime})$, where $V$ is the Green function of the fractional Laplacian in two-dimensions. 
In the case of an infinite domain $\mathcal{D}$, this Green function is a Riesz potential $V\sim r^{\alpha-2}$ with $r=|\mathbf{r}-\mathbf{r}^{\prime}|$ except when $\alpha$ is even, in which case $V\sim r^{\alpha-2}\left(\log r+C\right)$~\cite{Riesz,GreenFunctionGeneralizedLaplacian}.
Whatever $\alpha>0$, interactions between fluid particles are always long-range. 
We present in this paper numerical simulations of the freely evolving {\em{Galerkin-truncated}} dynamics of these 2D flow models, which is obtained by projecting Eq.~(\ref{eq:continuous_dynamics}) on the wave-numbers $\left| k \right|\le k_{max}$ and $\left| \ell \right | \le k_{max}$, where $k_{max}$ is the wave-number cut-off. 
This corresponds to an effective spatial resolution $N^{1/2}\times N^{1/2}$ with $N^{1/2}=3k_{max}$~
\footnote{We employ an energy-enstrophy conserving pseudo-spectral code with dealiasing using the Orzag's 2/3 rule.
Time integration were performed using Runge-Kutta scheme at order 4. 
We checked that energy and enstrophy decreased by less than $0.01\%$ for all runs presented in this letter.}. 
The initial $\alpha$-vorticity field is the same for all numerical experiments presented in this paper (see Fig.~\ref{fig1}).
 {It is characterised by a double peaked global distribution of $\alpha$-vorticity, and by a typical injection scale $k_i = 4$. 
Anticipating that the equilibrium state is always self-organized at the domain scale, choosing  $k_i \ge 2$ ensures that the initial condition is far from equilibrium.
This initial condition is "typical", in the sense that other initial conditions with similar injection length scale and similar global distribution yield similar results. 
%
}  

 {The initial eddy turnover time can be estimated as  $ t_i = (k_i / 2\pi)^{\alpha/2-2}  (2\pi /E^{1/2})$, and the time unit will be chosen for each numerical experiment such that $t_i=1$. 
The vorticity field is stirred by the turbulent flow during a few eddy turnover times with concomitant self-organization at domain scale and direct enstrophy cascade, until it reaches a QSS around  $t \sim 10$, which can for instance be quantified by checking that the isotropic energy spectrum does not change significantly over few dozens of eddy turnover times beyond that time.} 
Two striking features of the QSS are sumarized in Fig.~\ref{fig1}. 
First, there is a scale separation in space between erratic small scale fluctuations and a well-defined large scale flow structure organized at the domain scale.
%
%
Second, the large scale flow structure is drastically different depending on the value of $\alpha$.
 {We ask in the following whether equilibrium statistical mechanics of the truncated system and of the continuous system can account for those features.}
%
%

The truncated dynamics is a dynamical system with a finite number of degrees of freedom given by the Fourier components of $q$, for which a detailed Liouville theorem holds~\cite{Kraichnan_Motgommery_1980_Reports_Progress_Physics}.
This allows for a direct application of the equilibrium statistical mechanics machinery. 
Among the infinite number of conserved quantities by the continuous dynamics, only the $\alpha$-energy $E=\sum_{\mathbf{k}} E_{\mathbf{k}}$ and the $\alpha$-enstrophy $Z=\sum_{\mathbf{k}} Z_{\mathbf{k}}$ are conserved by Galerkin-truncated models, where  $E_{\mathbf{k}}=-q_{\mathbf{k}}\psi^*_{\mathbf{k}}$  and and  $Z_{\mathbf{k}}=|\mathbf{k}|^{\alpha}E_{\mathbf{k}}$ are the energy and the enstrophy of mode $\mathbf{k}$, respectively. 
Computation of equilibrium states of the truncated system in the thermodynamic limit ($N \rightarrow + \infty$) is a classical result predicting condensation of the $\alpha$-energy in the lowest-wavenumber mode ($\sum_{|\mathbf{k}|=1}E_\mathbf{k}=E$) and  a concomitant loss of  $\alpha$-enstrophy towards small scales \cite{kraichnan1967inertial,Kraichnan_Motgommery_1980_Reports_Progress_Physics}.
More precisely, for large $N$, Fourier modes  other than the lowest-wavenumber one have a contribution to the equilibrium state given by $<E_{\mathbf{k}}>=\frac{1}{4N}\left(\frac{Z-E}{|\mathbf{k}|^{\alpha}-1}\right)$,  where $\left<\cdot\right>$ stands for a temporal average, which shows equipartition of the enstrophy  $Z-E$ among the Fourier modes for sufficiently large $|\mathbf{k}|$~\cite{BasdevantSadourny}.

\begin{figure}

\includegraphics[width=\columnwidth]{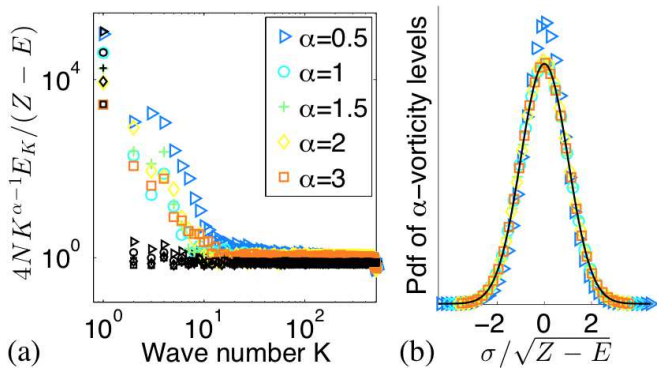}

\protect\caption{(Color online) a) Isotropic energy spectrum $E_K(K)$  normalized by $(Z-E)/(4NK^{\alpha-1})$ at large time ($t \approx 40$) for numerical simulations at resolution $N=1024^2$.  b) Corresponding local distributions of $\alpha$-vorticity levels. The black  symbols in panel (a) and the black line in panel (b) are  statistical mechanics predictions for the truncated system, without fitting parameters.}
\label{fig2}
\end{figure}

The ``isotropic energy spectrum''  defined as $E(K)=\left(\sum_{|\mathbf{k}|=K}E_{\mathbf{k}}\right) / \left(2\pi K\right)$ is shown in Fig.~\ref{fig2}-a for various values of $\alpha$.
The spectra are computed for only one snapshot at large time ($t \approx 40$), and we checked that averaging over many snapshots did not make any difference. 
As predicted by the theory, the energy is mostly condensed at the domain scale ($K=1$), and Fourier modes are thermalized with equipartition of $\alpha$-enstrophy into Fourier modes at large $K$, confirming previous numerical studies performed in the context of 2D Euler dynamics \cite{Kraichnan_Motgommery_1980_Reports_Progress_Physics,BasdevantSadourny,krstulovic2010galerkin} or surface quasi-geostrophic dynamics  \cite{TeitelbaumMininni}. 

The presence of a large scale flow containing most of the energy coexisting with wild small scale fluctuations of $\alpha$-vorticity  gives a strong incentive for a mean-field theory that would predict the  probability density  field $\rho(\mathbf{r},\sigma)$ to measure the $\alpha$-vorticity level $q=\sigma$ in the vicinity of point $\mathbf{r}$, and this is what is predicted by the MRS equilibrium statistical mechanics  \cite{miller1992statistical,SommeriaRobert:1991_JFM_meca_Stat}.
In this framework, all conserved quantities of the continuous dynamics can be expresssed in term of $\rho(\mathbf{r},\sigma)$. 
One can then count the number of microscopic configurations associated with each macroscopic state $\rho(\mathbf{r},\sigma)$ and compute the most probable macrostate $\rho(\mathbf{r},\sigma)$ satisfying the constraints of the problem. 
The theory predicts a concentration of all microscopic configurations close to the most probable macrostate $\rho(\mathbf{r},\sigma)$~\cite{michelrobert}.
The large scale flow is then given by  $\overline{q}(\mathbf{r})=\int\mathrm{d}\sigma\ \sigma\rho$, and the theory predicts a monotonic relation between $\overline{q}$ and $\psi$~\cite{miller1992statistical,SommeriaRobert:1991_JFM_meca_Stat}.
 {In  practice, following a standard procedure, the observed macrostate $\overline{q}$ will be computed through a local coarse-graining of the microscopic field $q$, by using a smoothing  operator with a typical length scale much larger than the effective grid mesh, but much smaller than the domain scale.} 
Equilibrium states of the truncated system  in the thermodynamic limit ($N \rightarrow +\infty$) are a subclass of MRS equilibrium states, characterized by  a linear $\overline{q}-\psi$ relation and by local gaussian fluctuations of $\alpha$-vorticity $\rho(\mathbf{r},\sigma)$ with variance $\left(Z- E\right)$~\cite{BouchetCorvellec2010,NasoChavanisDubrulle2010}.  
By contrast, when additional invariants than the energy and the enstrophy are taken into account, $\rho(\mathbf{r},\sigma)$ is in general non-gaussian. 
For instance, the initial condition shown on Fig.~\ref{fig1} has been constructed in such a way that the global distribution is close to a double peaked distribution, and MRS  theory  predicts in that case that the local probability distributions of the equilibrium states of the continuous dynamics should also be  double peaked distributions~\cite{BouchetVenaillePR12}.

\begin{figure}
\includegraphics[width=\columnwidth]{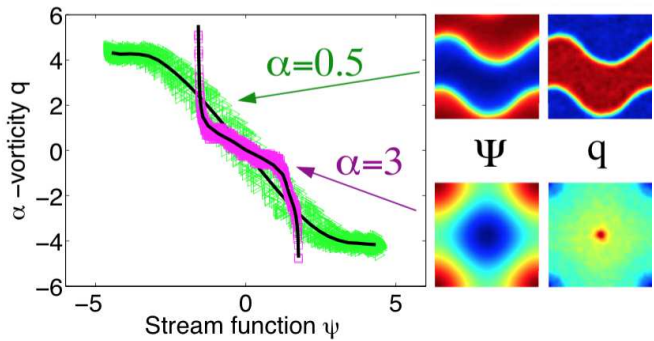} 
\caption{(Color online) $\overline{q}-\psi$ relation for the fields shown in Fig.~\ref{fig1} at $t\sim 100$. 
The plain black line is obtained by considering the averaged $\alpha$-vorticity value along a given streamline. }
 \label{fig3}
\end{figure}

%
%
%
To the best of our knowledge, we show in Fig.~\ref{fig2}-b the first observation of such local gaussian fluctuations for the vorticity field in numerical simulations of the Galerkin-truncated dynamics~\footnote{The pdf shown on Fig.~\ref{fig2}-b  are obtained by computing histograms of vorticity fluctuations along a  given contour of stream function, and summing over many contours. We checked that there were no significant variation of the distribution from one contour to another (expected for $\alpha<1$), and that distributions obtained in the vicinity of a given point lead to similar results.}, confirming theoretical predictions~\cite{BouchetCorvellec2010,NasoChavanisDubrulle2010}.
However, the success of the statistical mechanics theory of the truncated system is restricted to small scales:  the theory underestimates the contribution of intermediate wavenumbers $1<K<30$ in the spectra of Fig.~\ref{fig2}-a.
In addition, Fig.~\ref{fig3} clearly shows that the $\overline{q}-\psi$ relation of the large scale flow is non-linear: in the case $\alpha=0.5$, it has a $\tanh$-like shape, while it has a $\sinh$-like shape in the case $\alpha=3$.
In any case, such relations are not predicted by the statistical mechanics of the truncated system, which fails to account for the different large scale flow structures observed in Fig.~\ref{fig1}.

For any monotonic functional $\overline{q}-\psi$ relation, there exists at least one set of constraints such that the MRS equilibrium state associated with these constraints is characterized  by this functional relation \cite{bouchet2008}. 
In that respect,  the observed large scale flow is close to an equilibrium state of the continuous dynamics.
However, we see in Fig.~\ref{fig3} that fluctuations around the  observed $\overline{q}-\psi$ functional relation are present, and we checked that these fluctuations are independent of the wavenumber cut-off. 
This means that the large scale flow is not exactly stationary.
In addition, given our choice of an initial condition  characterized  by a global distribution of $\alpha$-vorticity levels with a double-peaked shape, MRS theory predicts that the  equilibrium state should be characterized by a $\tanh$-like shape, whatever the value of $\alpha$, see e.g. \cite{BouchetVenaillePR12}.
This means that the large scale flow obtained in the case $\alpha=0.5$ is close to the actual equilibrium state, but not  in the case $\alpha=3$.

\begin{figure}
\includegraphics[width=\columnwidth]{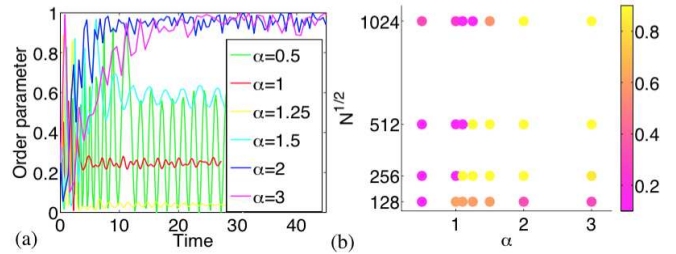}
\caption{(Color online) a) Temporal evolution of the order parameter $\mathcal{O}$ (see text) for various values of $\alpha$, in the run with resolution $1024^2$ b) Temporal average of the order parameter $\mathcal{O}$  for different values of $\alpha$ and wavenumber cut-off $k_{max}=N^{1/2}/3 $.}
\label{fig4}
\end{figure}

A transition from a unidirectional flow to a dipolar flow is expected when the $\overline{q}-\psi$ relation changes from a $\tanh$-like shape to a $\sinh$-like shape \cite{BouchetSimonnet2009}, and the large time flow structure  shown in Fig.~\ref{fig3} are consistent with these predictions.  
In order to study more quantitatively the transition from one state to the other when $\alpha$ is varied, it is useful to introduce an empirical order parameter  $\mathcal{O}=\frac{\min\left\{E_{(1,0)};E_{(0,1)}\right\}}{\max\left\{E_{(1,0)};E_{(0,1)}\right\}}$ comparing the energy of the lowest-wavenumber modes $(k,l)\in \left\{\left(1,0\right), \left(0,1\right)\right\}$. 
This parameter is zero in the case of a unidirectional state, and one in the case of a purely dipolar state. 
 {Fig.~\ref{fig4}-a shows the temporal evolution of $\mathcal{O}$ in the high resolution runs $N=1024^2$.  The flow is trapped in the dipole state when  $\alpha\ge2$. 
The order parameter decreases with $\alpha$  for $1.25<\alpha<2$, which shows the existence of a smooth transition from the dipole state to the unidirectional state. 
For $\alpha \approx 1$ the order parameter is characterized by small periodic or quasi-periodic oscillations, which are related to the presence of unmixed vortices. 
Such oscillations have previously been reported in the context of 2D Euler dynamics, for particular initial conditions~\cite{SegreKida,morita2011collective,dritschel2015two}.
For even smaller values of $\alpha$ (here $\alpha=0.5$), the unidirectional state is  characterized by periodic oscillations corresponding to large scale oscillations of the interface between two homogeneous regions of potential vorticity.
These states were previously observed for some particular initial conditions in the context of the 2D Euler equation~\cite{SegreKida}. 
 {In the presence of finite small scale dissipation, these oscillations were found to decay at sufficiently large time~\cite{YinMontgomeryClercx}; here effective dissipation occurs due to finite resolution and we observed that increasing resolution leads to an increase of the time scale for the decay of these oscillations.}
Fig.~\ref{fig4}-b shows  temporal averages of the order parameter $\mathcal{O}$  as function of $\alpha$ and resolution $N^{1/2}$.
Time average is performed over few dozens of eddy turnover times once the QSS is reached.  
We see a sharp transition from one regime to the other around $\alpha=1$ for low resolution runs, but the transition becomes smoother for higher resolution runs, as in Fig. 4-a. 
We checked that the dipolar state was robust when further increasing the resolution for $\alpha=2$.
We also found that decreasing the injection length scale $k_i$ shifted the transition from dipolar to unidirectional states below $\alpha=2$, but did not change qualitatively the results.} 
We show in the following that phenomenological arguments in limiting cases allows us to rationalize these observations.

In the limit $\alpha \rightarrow 0$, the  $\alpha$-vorticity can be written at lowest order as $q=\alpha\mathcal{L}\left[\psi\right]-\psi$, with $\mathcal{L}$ a  negative definite operator whose eigenmodes are Laplacian eigenmodes and whose  eigenvalues are increasing functions of  $|\mathbf{k}|$.
This is reminiscent of the $1\sfrac{1}{2}$ layer quasi-geostrophic model, another 2D flow model for which the advected tracer is $q=\Delta\psi-\psi/R^2$. 
 In the limit $R \rightarrow 0$, this flow model is known to spontaneously form regions where $q$ is homogenized, separated by sharp jets~\cite{arbicribbon,mcintyre}, which is expected either from cascade phenomenology~\cite{venaille2014ribbon} or from statistical mechanics arguments~\cite{bouchet2002emergence}. 
The formal analogy between $1\sfrac{1}{2}$ layer quasi-geostrophic model and 2D $\alpha$-turbulence in the limit $\alpha\rightarrow 0$ explains therefore the coarsening process resulting in the phase separation observed in Fig.~\ref{fig1} for  $\alpha = 0.5$.
Once the two regions of homogenized $\alpha$-vorticity are formed, their interface supports the existence of neutral (or Kelvin) modes oscillating periodically~\cite{batchelor2000introduction,dritschel1998persistence}. 
This prevents a complete relaxation towards the actual equilibrium state, which is characterized by a minimal  interface length~\cite{bouchet2002emergence}.
%

%

%
When $\alpha\rightarrow+\infty$, the stream function field is  dominated by the lowest-wavenumber modes $|\mathbf{k}|=1$ as soon as the $\alpha$-vorticity projections on the lowest-wavenumber modes ($q_{1,0}$ or  $q_{0,1}$) are non-zero.
The $\alpha$-vorticity field $q$ is sheared by this large scale flow, except at the two points where straining vanishes (provided that both $q_{1,0}$ and $q_{0,1}$ are non-zero).
Since the large scale flow is initially non stationary, the early evolution of the $\alpha$-vorticity field looks like chaotic mixing of a passive tracer due to a large scale flow (see Fig.~\ref{fig1} for $\alpha=3$).
This process leads to a background of homogenized $\alpha$-vorticity field, with two isolated blobs of $\alpha$-vorticity in the vicinity of the two points  where straining vanishes. 
Following this phenomenology, irreversible mixing in physical space through filamentation due to stretching prevents efficient mixing in phase space. 

 {For intermediate values of $\alpha$, a useful non dimensional parameter of the problem is given by $P=\left(\frac{2\pi}{L}\right)^{\alpha} \frac{E}{Z}$ comparing the enstrophy of the coarse-grained large scale flow with the total  enstrophy. 
This parameter varies from $0.5$ to $0.02$ in our simulations.  
It has been observed in the Euler case that an unstable initial condition converges in general towards a state close to the equilibrium one when $P$ is close to one~\cite{tabeling2002two}, while the dynamics relaxes towards a dipolar flow with two isolated vorticity peaks and  a background of homogenized vorticity  when $P$ is small~\cite{carnevale1992rates,Pomeau94}.
A qualitative reason for the failure of statistical mechanics prediction in that case is that excess enstrophy $Z-E\left(\frac{2 \pi}{L}\right)^{\alpha}$ initially at scale $2\pi / k_i$  contains most of the information on the large scale distribution of $\alpha$-vorticity, and yet does not contribute significantly to the dynamics of the large scale flow since it is rapidly lost into smaller scales through a direct enstrophy cascade.
In our numerical experiments  $Z$ is prescribed independently from $\alpha$, and the energy is $E\sim k_i^{-\alpha }Z $, which yields  $P\sim (k_i L)^{-\alpha}$.
We see that $P$ decreases when either $k_i$ increases or when $\alpha$ increases: as far as the convergence towards the dipolar state is concerned, changing the range of interactions has the same effect as changing the initial condition. 
We also see that $P$ tends to one when $\alpha$ tends to zero, which suggests that the dynamics is more likely to reach an equilibrium state in this limit, consistently with our numerical results.} 
%
%

\paragraph{Conclusion.} 

 {We have studied the dynamics of truncated two-dimensional flows in which the energy and the enstrophy were injected at a scale $2\pi / k_i$ smaller than the domain scale $L$, ensuring that the initial state is far from equilibrium.
We have shown that small scales features of the QSS following the initial  violent relaxation  are well described by the equilibrium statistical mechanics of the truncated system, but that the  corresponding large scale flow remains close to a stable state of the continuous dynamics whose topology depends strongly on the range of interactions, through the parameter $\left(\frac{2\pi}{L}\right)^\alpha \frac{E}{Z}$. 
When this parameter is small, which occurs whenever $\alpha$ is sufficiently large, the dynamics systematically relaxes towards a dipolar state through a filamentation. 
When this parameter is close to one, which occurs whenever $\alpha$ is sufficiently small, the system relaxes close to the equilibrium state of the continuous dynamics predicted by MRS theory through a coarsening process, but we observed persistent large scale oscillations preventing a complete relaxation.
On the basis of those results, we conjecture that weak long-range interacting systems are more prone to relax towards equilibrium  than strong long-range interacting systems, but that the time scale for complete relaxation may diverge as the range of interactions get weaker.
We focused in this paper on a small range of $\alpha$ that included existing physical models. Exploring a larger range of $\alpha$ will be needed to test in more detail the phenomenological argument obtained  in limiting cases.}

\begin{acknowledgments}
SR and AV acknowledge financial support from ANR-10-CEXC-010-01. AV acknowledges financial support from ERC-FP7/2007-2013-616811 and warmly thanks Isaac Held for suggesting the study of alpha-turbulence models in the scope of statistical mechanics approaches. 
\end{acknowledgments}

\bibliographystyle{phaip}
\bibliography{thermalisation}

\end{document}